\documentclass[12pt]{article}
\usepackage{esfconf}

\begin{document}

\def\AP#1#2#3{Ann. Phys. {\bf#1} (#2) #3}
\def\CMP#1#2#3{Commun. Math. Phys. {\bf#1} (#2) #3}
\def\CQG#1#2#3{Class. Quant. Grav. {\bf#1} (#2) #3}
\def\IJMPA#1#2#3{Int. J. Mod. Phys. {\bf A}{\bf#1} (#2) #3}
\def\JGP#1#2#3{J. Geom. Phys. {\bf#1} (#2) #3}
\def\JMP#1#2#3{J. Math. Phys. {\bf#1} (#2) #3}
\def\JPA#1#2#3{J. Phys. {\bf A}{\bf#1} (#2) #3}
\def\MPLA#1#2#3{Mod. Phys. Lett. {\bf A}{\bf#1} (#2) #3}
\def\PLB#1#2#3{Phys. Lett. {\bf B}{\bf#1} (#2) #3}
\def\NPB#1#2#3{Nucl. Phys. {\bf B}{\bf#1} (#2) #3}
\def\PRD#1#2#3{Phys. Rev. {\bf D}{\bf#1} (#2) #3}
\def\PRL#1#2#3{Phys. Rev. Lett. {\bf#1} (#2) #3}
\def\ZP#1#2#3{Z. Phys. {\bf#1} (#2) #3}

\def\hepth#1{{\tt hep-th/#1}}
\def\JHEP#1#2#3{J. High Energy Phys. {bf#1} (#2) #3}
\def\ss{\scriptstyle}
\def\sss{\scriptscriptstyle}
\def\*{\partial}
\def\punkt{\,\,.}
\def\komma{\,\,,}
\def\minus{\!-\!}
\def\+{\!+\!}
\def\={\!=\!}
\def\small#1{{\hbox{$#1$}}}
\def\half{\small{1\over2}}
\def\fraction#1{\small{1\over#1}}
\def\fr{\fraction}
\def\Fraction#1#2{\small{#1\over#2}}
\def\Fr{\Fraction}
\def\tr{\hbox{\rm tr}}
\def\eg{{\it e.g.}}
\def\ie{{\it i.e.}}
\def\etal{et al\.}

\def\w{\!\wedge\!}
\def\id{1\hskip-3.5pt 1}

\def\nl{\hfill\break\indent}
\def\nlni{\hfill\break}

\def\first{{1}${}^{st}$}
\def\second{{2}${}^{nd}$}
\def\third{{3}${}^{rd}$}

\def\a{\alpha}
\def\b{\beta}
\def\d{\delta}
\def\e{\varepsilon}
\def\c{\gamma}
\def\l{\lambda}
\def\ld{\lambda^\dagger}
\def\p{\psi}
\def\pd{\psi^\dagger}
\def\r{\varrho}
\def\s{\sigma}
\def\w{\omega}
\def\G{\Gamma}
\def\tG{\tilde\Gamma}
\def\S{\Sigma}
\def\tS{\tilde\Sigma}
\def\Z{{\Bbb Z}}
\def\C{{\Bbb C}}
\def\R{{\Bbb R}}
\def\H{{\Bbb H}}
\def\L{{\cal L}}
\def\M{{\cal M}}
\def\Ham{{\cal H}}
\def\ad{\hbox{ad}\,}
\def\Int{\int\limits}
\def\bra{\,<\!\!}
\def\ket{\!\!>\,}
\def\ra{\rightarrow}
\def\cross{\!\times\!}
\def\Tr{\hbox{Tr}\,}
\def\ker{\hbox{ker}\,}
\def\Re{\hbox{Re}\,}

\def\O{\Omega}
\def\o{\omega}
\def\tX{\tilde X}
\def\tS{\tilde S}
\def\tT{\tilde T}
\def\tY{\tilde Y}
\def\tZ{\tilde Z}

\def\ggr{\!\cdot\!}


\title{Generalised  11-dimensional supergravity}


\authors{Martin Cederwall, Ulf Gran, Mikkel Nielsen
and Bengt EW Nilsson}


\addresses{\centerline{Institute for Theoretical Physics}
\centerline{G\"oteborg University and Chalmers University of Technology}
\centerline{SE 412 96 G\"oteborg, Sweden}}


\maketitle


\begin{abstract}
The low-energy 
effective dynamics of M-theory, 
eleven-dimen\-sional supergravity, is taken off-shell in a manifestly 
supersymmetric superspace formulation. We show that a previously proposed
relaxation of the torsion constraints can indeed accomodate a current
supermultiplet. We comment on
the relation and application of this completely general formalism 
to higher-derivative ($R^4$) corrections.
This talk was presented by Bengt EW Nilsson at the Triangle Meeting 2000
``Non-perturbative Methods in Field and String Theory'' , 
NORDITA, Copenhagen, June 19-22, 2000, and by Martin Cederwall at the 
International Conference 
``Quantization, Gauge Theory and Strings'' in memory of Efim Fradkin,
Moscow, June 5-10, 2000.
The results presented in this talk are published in 
\cite{1}.
\end{abstract}



\section{Introduction}

One approach to probing M-theory 
at short distances is to consider the
effective action beyond its lowest order approximation given by the
second order (in $\#\hbox{(derivatives)}+\half\#\hbox{(fermions)}$)
action \cite{2}
\begin{equation}
\matrix{
-2\kappa^2S&=&\int d^{11}x\sqrt{-g}\left(R+\fr{2\cdot4!}H^{mnpq}H_{mnpq}\right)
+\fr6\int C\wedge H\wedge H \cr
&&\cr
&&+\hbox{\it terms with fermions}\komma\hfill\cr
}
\label{lagrangian}
\end{equation}
and investigate the higher-derivative corrections generated by the
microscopic theory.  Such  corrections at order $R^2$ and $R^4$ have been
extensively discussed in the literature, primarily in the context of
string theory and ten-dimensional effective actions, but also in the
eleven-dimensional context relevant to M-theory. The existence of
these terms can be inferred by a variety of means in string theory,
while in M-theory one must rely on anomaly cancellation arguments 
\cite{3,4}, or 
(superparticle) loop calculations 
\cite{5,6,7,8,9} 
together with the connection to string
theory and its effective action via dimensional reduction.

The methods used so far to deduce the existence of \eg\ $R^4$ terms in
eleven dimensions produce only isolated terms out of a large number of
terms making up the complete superinvariant that it belongs to. It is
of interest to have a better understanding of these superinvariants,
and there has consequently been a lot of work invested into the
supersymmetrisation of the isolated terms. In particular, $R^2$ and
$R^4$ terms in ten dimensions were considered already some time ago,
see ref. \cite{10} and references therein.
More recently also the $R^4$ term in eleven dimensions has been
investigated \cite{11} including a detailed study of superinvariants.

For these purposes it would be interesting to develop 
methods \cite{12} based on
superspace in eleven dimensions \cite{13} that would incorporate
supersymmetry in a manifest way. Although not yet developed into an
easily applicable formalism, $N$=1 supergravity in ten dimensions has been
constructed off-shell in terms of a linearised superspace lagrangian
\cite{14},
including some superinvariants \cite{15,16}, 
and should in principle lend itself to a
complete analysis of superinvariants and deduction of the
corresponding higher-derivative terms in ordinary component
language. The situation in eleven-dimensional M-theory is, however,
completely different due to the fact that an off-shell lagrangian
formulation with a finite number of auxiliary fields is not known and may 
not even
exist. From a general counting argument by Siegel and Ro\v cek
\cite{17} we know that this is true for $N$=4 super-Yang--Mills 
in four dimensions (and consequently also in ten dimensions)
but that maximally supersymmetric supergravity passes the test. 
The analysis carried out here, when completed, will
provide an answer to the question whether there exists
an off-shell lagrangian formulation in eleven dimensions or not.
In this respect the approach advocated here is parallel to the
discussion of ten-dimensional super-Yang--Mills theory carried out in
refs. \cite{18} and \cite{19}, which does in fact prove
that an  off-shell lagrangian based on these superspace fields 
does not exist.

To implement the symmetries of any M-theory effective action in a
manifest way, we will here follow ref. \cite{12} and define the theory in
superspace by means of the superspace Bianchi identities (SSBIs), which
are integrability conditions when the theory is formulated in terms of
superspace field strengths. From these
we will derive consistency conditions on the form of
the field equations. The analysis of the SSBIs will depend on the
structure of certain components of the supertorsion, and one particular 
goal is to
find connections between the various possible superinvariants and
consistent expressions for the components of the supertorsion.  The
structure of these components, as \eg\ which components can be set to
zero under which conditions, will be clarified by our analysis. This
is an important result since the torsion components are a vital input
when proving $\kappa$-invariance for M2 and M5 branes coupled to
background supergravity \cite{20,21,22} 
and M-theory corrected versions of it. In
fact, one should compare to the situation in IIA and IIB string theory
and the coupling to D-branes \cite{23,24,25,26}.
 Here it has been established that there
are higher-derivative background field corrections also on the world-sheets 
of the branes, see \eg\ ref. \cite{27} and references therein.
The presence of such terms complicates the issue
of $\kappa$-invariance and it becomes crucial to know the exact
form of the supertorsion and to understand 
its relation to the corrections both in
target space and on the brane.

Another aspect of the higher-derivative corrections is that it is to a large
extent unclear how supersymmetry organises the infinite set of such
terms into infinite subsets unrelated by supersymmetry. From previous work
both in ten and eleven dimensions we know that adding one bosonic
$R^2$ or $R^4$ term generates an infinite set of other terms of
progressively higher order in number of derivatives. This is clear in
any on-shell theory, as discussed in detail in the type IIB and heterotic 
cases in \eg\
ref. \cite{28,29,30}. 

In this talk we will make use of the fact that any conceivable
M-theory correction to the field equations must be compatible with
supersymmetry and local Lorentz invariance. This is built into the
SSBIs \cite{31,12} 
which when solved (the meaning of which is explained below)
produce constraints on the supertorsion and other superfields that
must be fulfilled by the corrections.  As a first step we prove in
this talk that the relaxed on-shell torsion constraints, argued for in
ref. \cite{12}, are correct and do not lead to the field equations that
follow from (\ref{lagrangian}).  
This will done without specifying the auxiliary
fields in terms of physical fields, making it possible to use the Weyl
superspace introduced by Howe \cite{32} to simplify the analysis of
the standard on-shell theory. Once the auxiliary fields are related to
physical fields, the role of Weyl superspace must be reconsidered, since
the identification will involve a dimensionful parameter ($\a'^3$ for the
$R^4$ term). This will be done elsewhere.

\section{Relaxed torsion constraints and off-shell solution of
Bianchi identities}
One of the most 
important results proved in ref. \cite{32} is that
inserting the single constraint
\begin{equation}
{{T}_{\a\b}}^{c}=2{\G}_{\a\b}^{c}\komma
\end{equation}
used in the superspace construction of eleven-dimensional supergravity 
\cite{13},
into the SSBIs leads to the field equations corresponding to the lowest
order lagrangian (\ref{lagrangian}).  This constraint must therefore be relaxed in
such a way that the equations that then follow from the SSBIs are able to
accommodate any higher-derivative correction terms to the field
equations. In order to explain how this is done we need some
details of the Weyl superspace formalism. This superspace is
coordinatised by $z^{M}=(x^m,\theta^\mu)$ where $m$ enumerates the
11 bosonic and $\mu$ the 32 real fermionic coordinates
respectively. The tangent superspace has as structure group the
Lorentz group (not a superversion of it) times Weyl rescalings, and
hence one introduces a supervielbein ${E_{M}}^{A}(z)$ and a superconnection 
${{\O}_{MA}}^{B}(z)={{\o}_{MA}}^{B}(z)+{K_{MA}}^{B}(z)$,
where $\o_{MA}{}^B=(\o_{Ma}{}^b,{1\over4}(\G^a{}_b)_\a{}^\b\o_{Ma}{}^b)$ 
is the Lorentz part and
${K_{MA}}^{B}=(2K_{M}{\d_a}^b,{K_{M}}{\d_{\a}}^{\b})$ the Weyl
part, and the flat superindex $A=(a,\a)$ contains an SO(1,10) vector 
index $a$ and a (Majorana) spinor index $\a$. 
The corresponding super-two-form field strengths
are (suppressing the wedge
product symbol in the product of superforms) 
\begin{equation}
T^{A}=DE^A=dE^A+E^B{{\O}_B}^A\komma\quad
{R_A}^B=d{{\O}_A}^B+{\O_A}^C{\O_C}^B\komma
\label{superfieldstrengths}
\end{equation}
and they satisfy the SSBI
\begin{equation}
DT^A=E^B{R_B}^A\komma\quad
D{R_A}^B=0\komma
\end{equation}
of which only the first one will be used in this talk. 
 Note that no separate superfield
corresponding to the four-form field strength is introduced
since both its Bianchi identity and field equation will emerge from
the analysis of the torsion SSBI. 

In order to obtain the form of the relaxed constraint given in
ref. \cite{12} we expand ${{T}_{\a\b}}^{c}$ in terms of irreducible tensors
by means of the basis for symmetric $\G$-matrices
$\G^{(1)}$, $\G^{(2)}$, $\G^{(5)}$, where $\G^{(n)}$ indicate a product of
$n$ antisymmetrised $\G$-matrices with weight one, \ie,
\begin{equation}
{{T}_{\a\b}}^{c}=2\left({{\G}_{\a\b}}^{d}{X_d}^c
+\fr2{{\G}_{\a\b}}^{d_1d_2}{X_{d_1d_2}}^c+
\fr{5!}{{\G}_{\a\b}}^{d_1\ldots d_5}{X_{d_1\ldots d_5}}^c\right)\komma
\label{aux}
\end{equation}
with the understanding that the $X$'s can be further decomposed into
irreducible tensors.  
In order to understand which parts of the tensors in eq. (\ref{aux}) that
are relevant, one needs to eliminate redundant superfields in a
systematic manner by imposing so called 
``conventional constraints''. The space does not allowed a detailed
discussion---we refer instead to ref. \cite{31}.
An analysis  of all the possible conventional 
constraints \cite{12} will leave 
only $X$'s in the 
representations 429 and 4290 in (\ref{aux}) as we will now discuss. 

Turning back to the components of the supertorsion at dimension 0 given in
eq. (\ref{aux}), we note that $X_d{}^c$ decomposes into representations of
dimension 65, with Dynkin label (20000), 55 with Dynkin label (01000), and
1 with Dynkin label (00000). Similarly, ${X_{d_1d_2}}^c$
goes into 11 (10000), 165 (00100) and 429 (11000), and 
${X_{d_1\ldots d_5}}^c$ into 330 (00010), 462
(00002) and 4290 (10002).
All antisymmetric tensors are set to zero.
At this point we
are left with the two fields which transform as 429 and 4290 under
SO(1,10). The way these appear in the supertorsion, \ie, in
${{T}_{\a\b}}^{c}$, suggests a close connection to the M2 and M5
brane, respectively, for the 429 and 4290. Although it seems easier to
deal with 429 we will in fact drop it and concentrate on the 4290
because of its probable relation to the anomaly canceling term related
to the M5 brane. (The field of interest with dimension 0, 
$\a'^3W^3+\ldots$ ($W$ is the Weyl tensor), 
does not contain the representation 429.) 
This will have to appear in the SSBI for the
four-form superfield strength which hence will read $d*H=\half
H^2+X_{(8)}$, where $X_{(8)}$ is the eight-form polynomial in the curvature
that was introduced in this context in ref. \cite{3,4}. 
 In this talk, however,
we will not take the analysis this far but instead show that the
relaxed torsion constraint \cite{12}
\begin{equation}
T_{\a\b}{}^{c}=2\left(\G_{\a\b}{}^{c}+
\fr{5!}\G_{\a\b}{}^{d_1\ldots d_5}X^{(4290)\,c}_{d_1\ldots d_5}\right)
\end{equation}
is general enough to lift the field equations coming from (\ref{lagrangian}). 


More details are found in \cite{1} and a more 
complete discussion will be presented elsewhere
\cite{34}. The method for solving the SSBI, $DT^A=E^B{R_B}^A$, 
is to extract its component equations and solve these by increasing
dimension.
The equation of lowest dimension, $\half$, is the one
multiplying the three-form $E^{\a}E^{\b}E^{\c}$ and with
$A=a$,
\begin{equation}
0={R_{(\a\b\c)}}^d=D_{(\a}{T_{\b\c)}}^d+{T_{(\a\b}}^E{T_{|E|\c)}}^d\komma
\label{dimhalfssbi}
\end{equation}
where $(\ldots)$ indicates symmetrisation of the indices (except for the
ones between bars $|\ldots|$). This equation can be decomposed into a
large number of equations, each one corresponding to an irreducible tensor
appearing in the decomposition of the symmetric product of three
spinors times a vector, which is the tensor structure of the SSBI
(\ref{dimhalfssbi}). When the expansions of the torsion 
components are inserted
into these irreducible tensor equations, all irreducible tensor parts
of the torsion will drop out except the ones that
coincide with the representation specifying the equation.

We should also mention that we restrict ourselves to a linearised analysis.
A more non-linear treatment is feasible, at least in the original 
fields, but here the ordinary supergravity fields and the auxiliary ones 
are treated on equal footing. In this talk we also neglect vector
derivatives on the auxiliary superfield $X^{4290}$.

Since the equation (\ref{dimhalfssbi}) involves the fields at $\theta$ level
in $X^{(4290)}$, we need to expand $D_{\a}{X_{a_1\ldots a_5}}^b$ as well as
the dimension 1/2 torsion components into
irreducible tensors. 
In fact, as a consequence of using Weyl superspace, with the
extra conventional constraints associated with the Weyl connection, we find
that the torsions involved in this SSBI are uniquely
determined by the components of $D_{\a}{X_{a_1\ldots a_5}}^b$. 
Thus if we set $X^{(4290)}$ to zero these
torsions will vanish without invoking any extra assumptions, a result
that also follows from the work of Howe in ref. \cite{32}. 
In this talk, we spare the reader from the exact expression for the 
torsion components in terms of $X^{(4290)}$. The calculation involves a 
certain degree of technical complexity, which is left for ref. \cite{34}. 

We now turn to the SSBIs with dimension 1. There are two such
equations, namely 
\begin{equation}
\matrix{
R_{\a\b c}{}^d=2D_{(\a}T_{\b)c}{}^d+D_cT_{\a\b}{}^d
	+T_{\a\b}{}^ET_{Ec}{}^d+2T_{c(\a}{}^ET_{|E|\b)}{}^d\komma\hfill\cr
R_{(\a\b\c)}{}^\d=D_{(\a}T_{\b\c)}{}^\d+T_{(\a\b}{}^ET_{|E|\c)}{}^\d\punkt
\hfill
\cr
}
\label{DimOneBI}
\end{equation}
To deal with these equations, we must expand the superfield $X^{(4290)}$
at the $\theta^2$ level, and take into account the results already
obtained at $\theta$ level. As mentioned above, we aim at showing that the
introduction of the auxiliary fields generates a right-hand side of the
spinor part of the equation of motion for the gravitino field. 
At present, we therefore only need to consider the irreducible tensors
at dimension 1 whose spinorial derivative contains a spinor.
These are all forms, with rank from zero to five. We state the
two- and three-forms, which are the ones that will eventually survive:
\begin{equation}
\matrix{
T_{a\b}{}^{\c}&=&\fr6(\G^{d_1d_2d_3})_\b{}^\c A_{d_1d_2d_3a}
+\fr{24}(\G_a{}^{d_1\ldots d_4})_\b{}^\c A'_{d_1\ldots d_4}\cr
&&+\fr2(\G^{d_1d_2})_\b{}^\c A_{d_1d_2a}
+\fr6(\G_a{}^{d_1d_2d_3})_\b{}^\c A'_{d_1d_2d_3}\hfill\cr
&&+\ldots\hfill\cr}
\end{equation}
Since now the curvatures entering eq. (\ref{DimOneBI}) 
are non-zero taking values in
the structure group, they must be eliminated. From the $R_{\a\b c}{}^d$
we get the information that the symmetric traceless part in $cd$ has to vanish,
and the rest are used to eliminate $R_{\a\b\c}{}^\d$ by the structure
group condition. In contrast to the equations at dimension $\half$, where
the full representation content of the index structure of the SSBI made
impact on the fields (to the extent that the representations were present
at level $\theta$ in $X^{(4290)}$), some equations now turn out to be
linearly dependent. A naive counting of fields and equations fails, and,
as we will see,
this is absolutely essential in order for the auxiliary superfield to contain
components entering the equations of motion.
This exceptional behaviour relies on the exact form of the 
solutions at dimension $\half$,
and comes at work for the three-forms, where three
equations reduce to two, and for the four-forms, where all three equations
are identical. 
The zero-, one-, two- and five-forms at second level in $X$
are set to zero (modulo terms $\sim D_bX_{a_1\ldots a_5}{}^b$ in
the five-forms), and the relevant surviving part is parametrised as
\begin{equation}
\matrix{
\fr{10}D_{[\a}D_{\b]}X_{a_1\ldots a_5,b}				
\hfill\cr
=\G_{[a_1a_2a_3}{}^eV_{a_4a_5]be}
	+\G_{b[a_1a_2}{}^eV_{a_3a_4a_5]e}
	-\Fr67\eta_{b[a_1}\G_{a_2a_3}{}^{e_1e_2}V_{a_4a_5]e_1e_2}	
\hfill\cr
+\G_{a_1\ldots a_5}{}^{e_1e_2e_3}W_{be_1e_2e_3}
	+\G_{b[a_1\ldots a_4}{}^{e_1e_2e_3}W_{a_5]e_1e_2e_3}
	-\Fr67\eta_{b[a_1}
		\G_{a_2\ldots a_5]}{}^{e_1\ldots e_4}W_{e_1\ldots e_4}	
\hfill\cr
+\G_{[a_1a_2a_3}V_{a_4a_5]b}
	+\G_{b[a_1a_2}V_{a_3a_4a_5]}
	-\Fr67\eta_{b[a_1}\G_{a_2a_3}{}^eV_{a_4a_5]e}			
\hfill\cr
+\G_{a_1\ldots a_5}{}^{e_1e_2}W_{be_1e_2}
	+\G_{b[a_1\ldots a_4}{}^{e_1e_2}W_{a_5]e_1e_2}
	-\Fr67\eta_{b[a_1}\G_{a_2\ldots a_5]}{}^{e_1e_2e_3}W_{e_1e_2e_3}\hfill
\cr
+\ldots\hfill\cr
}
\end{equation}
with the relations
\begin{equation}
\hbox{$A+2A'+{2\cdot5\over7\cdot11\cdot23}\left(547\,V+2^5\cdot3^3\cdot17\,W\right)=0$}
\end{equation}
for the four-forms, and
\begin{equation}
\matrix{
A=-{2^2\over3\cdot11\cdot23}\left(89\,V+2^2\cdot3\cdot5\cdot139\,W\right)\komma
\hfill\cr
A'={2^3\over3\cdot11\cdot23}\left(2\cdot47\,V-3\cdot5\cdot41\,W\right)\hfill\cr
}
\end{equation}
for the three-forms. The linear dependence already mentioned makes us
confident in these expressions. 
In the unmodified supergravity ($V=W=0$), one four-form 
in the dimension 1 torsion
survives, and is identified with the four-form field strength $H$.
In the present situation, it is a priori not obvious which combination
of the three surviving four-forms that should be identified with this
physical field (the criterion being that it is closed), and the answer to
this question will have to await the solution of the SSBIs at dimension 2.

One further result at dimension 1 is that the Weyl part of the curvature,
$G_{\a\b}=\fr{32}R_{\a\b\c}{}^\c$, vanishes, 
as was shown in ref. \cite{32} for
the unmodified supergravity. 
This is a very positive sign, since it
indicates that the theory even with the relaxed torsion constraints 
we have used to
take it off-shell is equivalent to one
with only the Lorentz group as structure group as discussed by Howe 
\cite{32}. We will comment on this further in 
the concluding section.

Finally, we consider the SSBIs at dimension $\Fr32$, which read
\begin{equation}
\matrix{
2R_{\a[bc]}{}^d=D_\a T_{bc}{}^d+2D_{[b}T_{c]\a}{}^d
	+2T_{\a[b}{}^ET_{|E|c]}{}^d+T_{bc}{}^ET_{E\a}{}^d\komma\hfill\cr
2R_{a(\b\c)}{}^\d=D_a T_{\b\c}{}^\d+2D_{(\b}T_{\c)a}{}^\d
	+2T_{a(\b}{}^ET_{|E|\c)}{}^\d+T_{\b\c}{}^ET_{Ea}{}^\d\punkt\hfill\cr
}
\end{equation}
In an unconstrained superfield in the representation 4290, there are
two spinors at level $\theta^3$. The index structure of the SSBIs at this
level also contains two spinor equations. We have to take into account what
has been learned about $X^{(4290)}$ at lower levels. Specifically, 
at dimension 1 some of the antisymmetric tensors containing a spinor at
the next level vanished. Miraculously, again, all of these go into the same
linear combination, while the spinor coming from the three- and four-forms
survives. One spinor thus remains, and goes into part of the field equation
for the Rarita--Schwinger field, which then reads
\begin{equation}
t_\a={17\over2^2\ggr3^3 \ggr 5^2 \ggr 7 \ggr 11 \ggr 13 \ggr 61}
(\G^{b_1b_2b_3}){}^{\b\c}(\G^{b_4b_5a}){}_\a{}^\d
	D_{[\b}D_\c D_{\d]}X_{b_1\ldots b_5,a}\komma
\end{equation}
where $t_\a$ is the spinor part of the decomposition of the dim-3/2 torsion
into irreducible representations:
$T_{ab}{}^{\c}=t_{ab}{}^{\c}+2(\G_{[a}t_{b]})^\c+(\G_{ab}t)^\c$.
Also the spinor component of the Weyl curvature, 
$G_\a=\fr{32}(\G^a)_\a{}^\b R_{a\b\c}{}^\c$, is set to zero.

These calculations involve some rather heavy $\Gamma$-matrix algebra which
has been facilitated enormously by the development of a Mathematica based 
program \cite{35}. In particular, the results in this talk rely on a large
number of Fierz identities which, as explained below, can be 
completely systematised. By using the algebraic program to compute some
small number of final coefficients, any computation requiring 
Fierzing is easily dealt with.

\section{Conclusions}
We have demonstrated that, through a series
of seemingly miraculous numerical coincidences (which, however, due to the 
similarities with ten dimensions \cite{36}, both regarding the constraints and the subsequent manipulations of the SSBIs, one would strongly expect to occur),
 the relaxation of the torsion constraint 
at dimension zero is capable of accommodating an off-shell formulation.
By off-shell we here simply mean that the equations of motion from 
(\ref{lagrangian}) are  
relaxed by the introduction of a current supermultiplet, contained in the
supertorsion along with the supergravity multiplet.
In this sense, the term ``on-any-shell'' might be more appropriate. 
However, since the non-existence of an off-shell action has to our
knowledge not been proven, the possibility is not ruled out that the auxiliary
fields produced by this formalism are the correct ones for the construction
of such an action.
We would like to stress that since the degrees of freedom contained in
the eleven-dimensional supergravity multiplet describe only low-energy
effective dynamics of M-theory, and this system is not supposed to be subject
to quantisation, the absence of an action at this level is completely
acceptable.
The results are sofar partial. We have not yet investigated all equations
of motion. In a following paper \cite{34}, 
we will give a more detailed account
of the calculations.

An obvious application of the formalism, as mentioned in the introduction, 
is to use it to derive higher-derivative corrections to M-theory, beginning
with $R^4$ terms and their superpartners. 
The identification of our auxiliary field $X^{(4290)}$ as a supergravity
self-interaction clearly breaks Weyl invariance. It is encouraging to note
that the corresponding curvatures vanish, as far as our analysis goes,
which indicates that the correct procedure is to restrict to the Lorentz
structure group in order to avoid ambiguities in the definition of
Weyl weights, while retaining the corresponding conventional constraints.

Brane dynamics in general backgrounds is most conveniently described in 
terms of quantities pulled back from target superspace to the world-volume.
It is known that $\kappa$-symmetry quite generally demands the background
fields to be on-shell. This must still be true for branes in backgrounds
modified by higher-derivative corrections. We believe that our formalism 
will be essential for such an analysis. One question arises directly: Is 
the action for \eg\ the M2-brane still given by the same expression,
\begin{equation}
S\sim\int d^3\sigma\sqrt{-g}+\int C\komma
\label{MTwoAction}
\end{equation}
so that the corrections come only through the pullbacks of the modified
background fields, or is this form changed?
We have not discussed the superspace tensor fields in this talk, but by 
analysing the dimension zero identity, one
realises that the equation $dH=0$ demands $H$ to have non-vanishing
components even at negative dimensions. These will appear in a 
$\kappa$-transformation of the WZ term in eq. (\ref{MTwoAction}), but do not have
any torsion counterpart to cancel. 
We hope to be able to come back also to this issue. 

Finally, it 
would also be interesting to investigate in a strict sense whether the
assumption of locality, which is implicit in our work, limits the current
multiplet to self-interactions of the supergravity multiplet or whether
there are traces of interactions with other M-theory states.

\vskip.5cm
\noindent{\bf Acknowledgements.}
Martin Cederwall is grateful to the organisers of the Efim Fradkin
Memorial Conference for their hospitality and helpfulness.
 The authors wish to thank Kasper Peeters, Pierre Vanhove and Anders
Westerberg for very inspiring discussions and for their generous
communication of work in progress \cite{11}. 
This work is partly  
supported by EU contract HPRN-CT-2000-00122 and by the Swedish Science
Research Council (NFR).
For some representation theoretical considerations, the program 
LiE \cite{37} has been useful.

\vfill\eject

\def\xit{\it}
\def\xrm{\rm}

\end{document}